\newcommand{\version}{November 22, 2026}
\documentclass{onecolartcl}

\title{\texorpdfstring{\begin{flushright}
		{\small LA-UR-26-24423}
	\end{flushright}\vspace{2em}}{}%
A new Fermi kinetic phase transition model for metals including hysteresis effects
}

\author{Ann E. Mattsson, Daniel N. Blaschke\texorpdfstring{\footnote{corresponding author} }{}, Michael B. Prime, \texorpdfstring{\\}{}David R. Jones, Saryu Fensin, and Abigail Hunter}

\date{\version}

\graphicspath{{figures}}

\begin{document}

\maketitle

\thispagestyle{empty}
\begin{center}
	\renewcommand{\thefootnote}{\fnsymbol{footnote}}
	\vspace{-0.3cm}
	Los Alamos National Laboratory, Los Alamos, NM, 87545, USA
	\\[0.5cm]
	\ttfamily{E-mail: aematts@lanl.gov, dblaschke@lanl.gov, prime@lanl.gov, \\djones@lanl.gov, saryuj@lanl.gov, ahunter@lanl.gov}
\end{center}

\begin{abstract}
We present a new phenomenological model for phase transformation (PT) kinetics in metals, the ``Fermi Kinetic Phase Transition (KPT) Model''.
It is designed such that it captures the main macroscopic features of our previously developed micro-structure dependent model, but at a fraction of the computational cost.
Using four model parameters, the Fermi KPT model performs better than other phenomenological PT kinetics models in the literature, as shown by our present comparisons to experimental data for iron and tin.
\end{abstract}

\setcounter{tocdepth}{2}
\tableofcontents

\section{Introduction}
\label{sec:intro}

Many crystals exhibit solid-solid phase transitions at either high pressure and/or temperature, where the crystal structure changes are also accompanied by a sudden change in volume \cite{Sikka:1982,Gornostyrev:1999,Davis:2007,Rigg:2009,Smith:2013,Zong:2014,Lazicki:2015,Gomez:2019,Barton:2022,Liu:2023,Yao:2024,Pei:2024,Li:2025,Hunter2026JMR}).
Equations of state typically capture sharp phase boundaries, which are appropriate for sufficiently slow-changing external conditions in temperature and pressure.
Under high rate loading, such as shock loading, for example, the time it takes for a phase transformation to occur becomes crucial in order to predict the extent of damage that can occur \cite{Resseguier:2008,Righi:2023}.
Furthermore, many studies (including some under quasi-static loading), have confirmed a pronounced hysteresis for pressure driven phase transitions \cite{Giles:1971,Sikka:1982,Taylor:1991,Rigg:2009,Zong:2014,Merkel:2020,Righi:2023}.

One example of a material that exhibits a hysteresis in its phase change behavior is iron, which has a  body-centered cubic (bcc) crystal structure ($\alpha$-iron) at room temperature and ambient pressure.
Per the phase diagram, at pressures of around 13 GPa its crystal structure changes to hexagonal-close-packed (hcp) $\epsilon$-iron (see e.g. \cite{Takahashi:1964,Barker:1974,Taylor:1991,Boettger:1997,Kalantar:2005,Kadau:2005,Jensen:2009,Bastea:2009,Hawreliak:2011,Smith:2013,PhysRevB.93.214108,Yao:2024}). 
However, when the material goes from $\alpha$ to $\epsilon$-iron via shock loading, it has been shown that the phase change occurs at pressures higher than 13 GPa \cite{Giles:1971,Merkel:2020,Smith:2013}.
This effect is evident for the reverse transition as well, which occurs at much lower pressures than expected, closer to 10 GPa \cite{Giles:1971,Taylor:1991,Merkel:2020}.
The variation in when these phase transitions occur versus the equilibrium value of 13 GPa produces the aforementioned hysteresis behavior and is due to the kinetics of the phase transitions.

Tin is another example of a material that exhibits solid-solid phase transformations and kinetics effects under high rate loading \cite{Davis:2007,Lazicki:2015}. 
At room temperature and ambient pressure, tin has a body-centered tetragonal (bct) crystal structure known as $\beta$-tin or `white' tin, which then transforms to 
a more compact bct
structure called $\gamma$-tin at high pressures ($\sim9.4$ GPa). 
The $\beta$ to $\gamma$-tin transition has also exhibited hysteresis in the forward/reverse transition on the order of 2-4 GPa \cite{PhysRevB.109.104116}.
Tin has other high pressure solid-solid phase transformations (e.g., bct-bcc and bcc-hcp) that have been observed via diamond anvil cell experiments, and the hysteresis in these transformations are even larger than that observed for the $\beta$ to $\gamma$ transition \cite{PhysRevB.109.104116}.

There have been several modeling approaches developed to capture and investigate phase transition kinetic effects, ranging from atomistic through phenomenological models. 
Atomistic methods have both the advantage and limitation of high resolution.
Advantageously, this allows the method to capture nucleation phenomena and the effect of defects, such as interstitials, dislocations, interfaces, etc., on phase transition behavior \cite{10.1063/5.0117234}. 
However, with this high resolution, the limitation becomes the accessible length and time scales.
Nevertheless, several atomistic studies have focused on the timescale or kinetics of solid-solid phase transformations occurring under high rate loading conditions \citep{Bertrand:2013,Zong:2014,Pang2014SR,Gunkelmann:2015,Guo:2021,Ma:2022,Daphalapurkar:2024}. 
It is worth noting that the accuracy of atomistic simulations depends upon the interatomic potential employed. 
While these approaches do not have parameters that need to be informed as is the case for meso- and macro-scale approaches, the interatomic potential does impact the overall results, including those for phase transformation kinetics \cite{Mueller_2007,SUIKER20132273}. 

Moving to mesoscale modeling approaches, traditionally, phase field models have been developed and applied to study phase transitions. A large body of literature is available on such approaches, the interested reader can refer to \cite{Elder:2007,Gomez:2019,Ansari:2021,Lahiri:2022,Tourret:2022,Yao:2024,Xu:2025} and references therein for a review.
In essence, phase field approaches simplify the interface structure, which enables accessibility to longer timescales and larger simulation sizes. 
While this reduction in resolution does neglect several atomic level features naturally accounted for in atomistic approaches, phase field approaches still remain tightly tied to the microphysics of the phase transformation process. 
For example, recent work presented a phase transformation kinetics model that generalizes the Levitas-Preston (LP) phase field model of martensite phase transformations to arbitrary pressures \cite{Blaschke:2025a}. 
This model accounts for nucleation on various microstructural features, e.g. dislocations and grain surfaces, edges, and vertices, in addition to using microstructural information to determine the nucleation rate of the new phase as well as the interface speed of the pressure driven phase transformation. 
This model is used later in this paper as a comparison point for more phenomenological models discussed. 

Informing mesoscale models such as phase field approaches can be non-trivial, and depending on the model can require many parameters.
Many are material parameters (e.g., lattice parameters, stiffness tensor, etc.) that are relatively straightforward to determine; however, many others connect to specific mechanisms that are difficult to determine experimentally and require atomistic approaches.
For example, models targeting phase transition kinetics could require information about interfacial energies and/or speed of the two-phase interface at a given pressure, which can potentially be computed with molecular dynamics simulations (e.g., \cite{Daphalapurkar:2024}).
Other microstructural information can be informed by detailed characterization such as dislocation density, average grain size and orientation distribution and grain boundary thickness. 
Understanding the dependence of results on each of these parameters along with the effect on uncertainties in these parameters is an area of open research and requires detailed simulation studies.

At the largest of length/time scales, phase transition kinetics theory aims to predict not only the hysteresis curve for a phase transition, but also how a phase transition evolves with time over very short time scales \cite{Greeff:2016}.
While more phenomenological, such formulations have been widely adopted for two main reasons: (1) computational speed which enables the study of component relevant length and time scales, and (2) relatively few unknown parameters than can be straightforwardly calibrated with experimental data.
Although, these parameters typically do not have direct relation to the microphysics of the phase transformation process nor microstructural features.
Perhaps on the extreme end, earlier phenomenological models for phase transition kinetics have very few model parameters, but then are challenged to capture all relevant physics particularly under wide ranges of loading conditions.
For example, the Greeff model \cite{Greeff:2016} has only two model parameters which, as we argue in this paper, cannot always be calibrated to experimental data, especially in high strain rate scenarios.
To address this challenge, we present a new phenomenological model, the ``Fermi Kinetic Phase Transition Model'', which is based on the Fermi-Dirac distribution and has four model parameters for every (direction-dependent) phase boundary.
We show that this model can capture phase transitions observed under shock loading in tin.
In addition, we present comparisons of this new model with the Greeff model \cite{Greeff:2016}, and the model discussed above which is microstructurally aware and that we label the `micro' model \cite{Blaschke:2025a}.

The outline of our paper is as follows:
We start by presenting the derivation of a new phenomenological model called the Fermi Kinetic Phase Transition (FKPT) model in Section \ref{sec:fermimodel}, which is meant to address shortcomings of the Greeff model detailed in Ref. \cite{Greeff:2016}.
Section \ref{sec:validate} is dedicated to validating this new model by comparing it to data of Smith et al. on iron \cite{Smith:2013} as well as recently measured gas-gun flyer-plate impact data on tin presented in Section \ref{sec:tingundata}.
As mentioned, we compare the Fermi model to two other models, the Greeff model reviewed in Section \ref{sec:greeff} and the micro-structure aware model --- or micro model --- presented in Ref. \cite{Blaschke:2025a}, which is reviewed in Section \ref{sec:modelreview}.
The analytic equation of state (EOS) for tin that we use, along with other simulation details are presented in subsequent Section \ref{sec:eos}.
In Section \ref{sec:comparison}, we highlight where each of the three models does well in capturing the essential features, concluding that the Fermi model performs best across all strain rate scales.

\section{The Fermi Kinetic Phase Transition (KPT) model}
\label{sec:fermimodel}
 
In a first take on a phenomenological model for phase transitions it is desirable to have one parameter describing the nucleation phase and one parameter giving the rate of the phase transition once it starts.
Aiming at having two fully decoupled fitting parameters for the nucleation and the transition rate, we propose to use the Fermi-Dirac distribution as a model for the mass fraction variation over a phase transition.
In our discussion below, we will further see that both parameters need to be strain rate dependent in order to capture all the features seen in the experimental data.

\subsection{Using the Fermi-Dirac distribution for rates in a phase transition}

\begin{figure}[ht]
\centering
\includegraphics[width=0.5\linewidth]{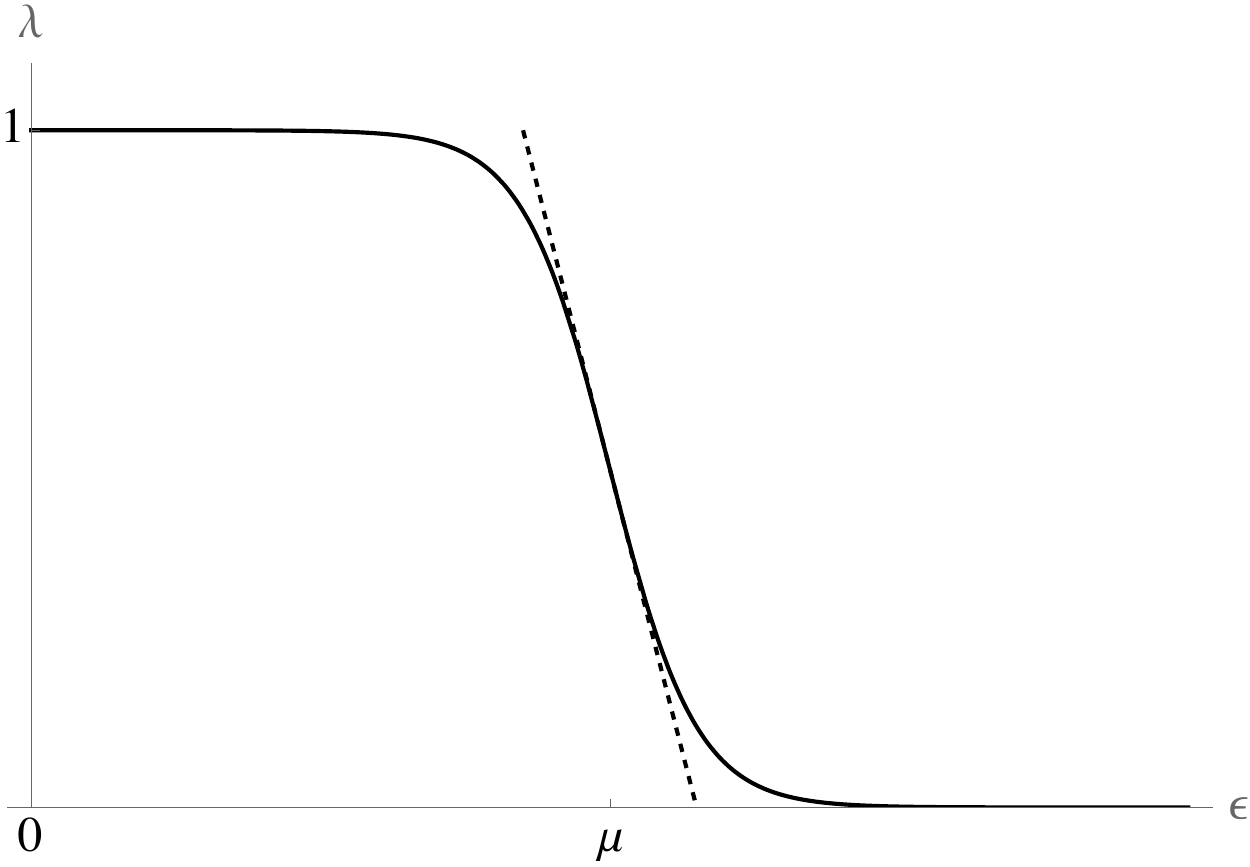}%
\includegraphics[width=0.5\linewidth]{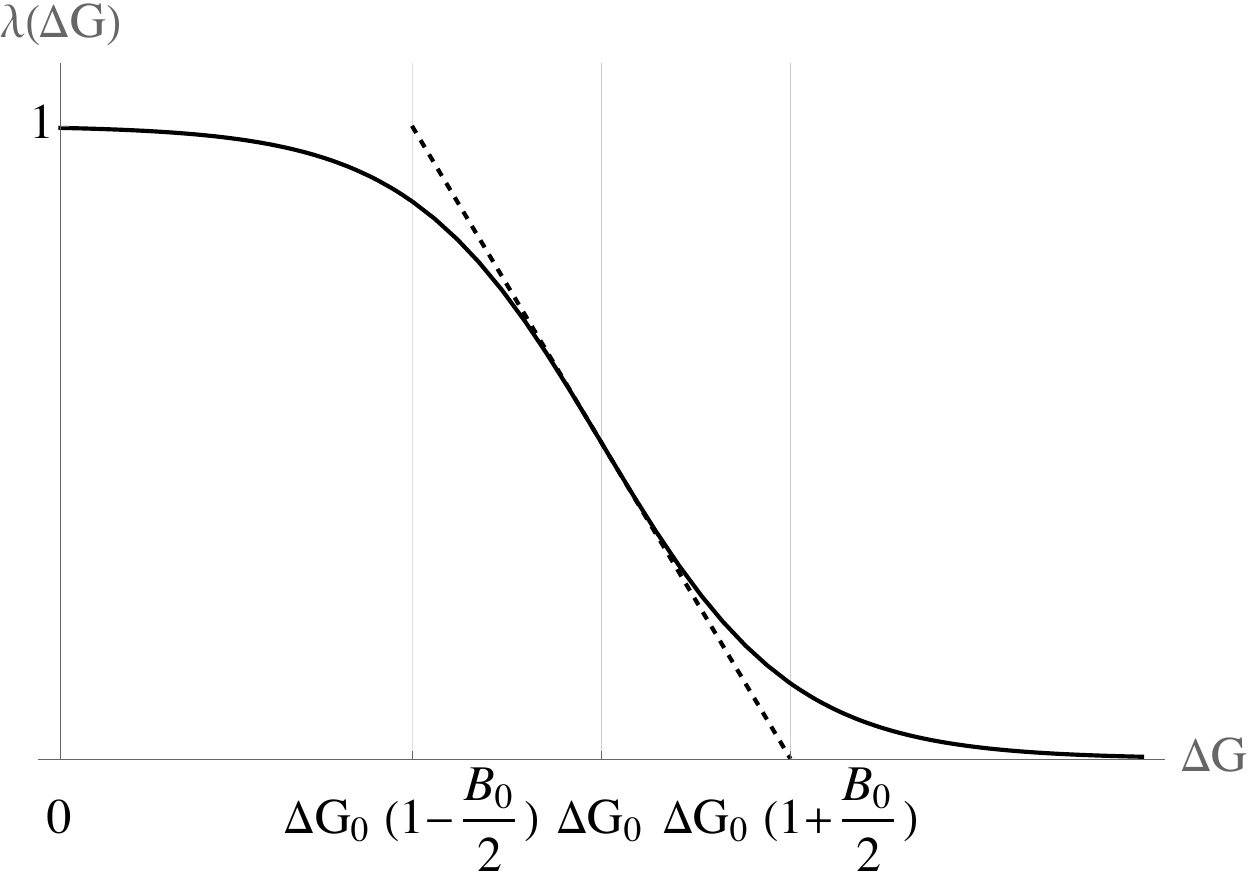}
\put(-0.75\columnwidth,0.341\columnwidth){\textbf{(a)}}
\put(-0.25\columnwidth,0.341\columnwidth){\textbf{(b)}}
\caption{On the left (a) we show the Fermi-Dirac distribution (full line) and its derivative at $\mu$ (dashed line).
On the right (b) we show the Fermi KPT model. $\Delta G_0$ is related to the nucleation energy barrier and sets the energy scale of the model. The width of the phase transition is $B_0$ in terms of this energy scale. Note that $\Delta G$ is positive, a negative $\Delta G$ will not trigger a phase transition.}
\label{fig:FKPT}
\end{figure}

The Fermi-Dirac distribution describes the electron occupancy of states when subject to a specific temperature, $T$ ($k_B$ is the Boltzmann constant):
\begin{equation}
    \lambda(\epsilon) = \frac{1}{e^\frac{\epsilon - \mu}{k_B T}+1} \, .
\end{equation}
At zero temperature, all states below the Fermi energy level, $\mu$, are occupied by one electron while no states above the Fermi level are occupied, thus, the distribution is a step function at $\mu$.
At higher temperatures the states around the Fermi level are partially occupied as is shown in Figure~\ref{fig:FKPT}.
The two parameters, $\mu$ and $T$, determine the form of the curve, with $\mu$ being the energy at which the level is half occupied and $T$ determining the slope at this energy, $\frac{1}{4 k_B T}$.

The curve in the left panel of Figure~\ref{fig:FKPT} could describe a phase transition. Replacing $\epsilon$ with the Gibbs Free energy difference, $\Delta G$, between the two phases, $\mu$ would represent the excess energy needed for the transition to start and be fully developed, that is, the nucleation energy barrier, $\Delta G_0$.
The slope of the curve 
determines how fast the transition is when fully developed, at $\Delta G_0$, and sets the width of the phase transition. Using the nucleation barrier as an energy scale, we write the mass fraction of the initial phase as
\begin{equation}
    \lambda_m(\Delta G) = \frac{1}{e^{4 \frac{\frac{\Delta G}{\Delta G_0} - 1}{B_0}}+1} \qquad \ \ \Delta G > 0 \, .
    \label{eqn:FKPT}
\end{equation}
This function is shown in the right panel of Figure~\ref{fig:FKPT} where the two parameters $\Delta G_0$ and $B_0$ are also explained. Note that with the scaling we use, the width, $B_0$, and the energy difference, $\Delta G_0$, where only half of the mass in the initial phase remains, can be set independently of each other. 

At this point we should, however, point out that in a real system, $\Delta G$ is seldom growing linearly beyond the very start of the phase transition and the resulting mass fraction profile as a function of \emph{time} is distorted accordingly.
The advantage of this Fermi KPT model is, however, obvious if we compare to the Greeff KPT model (see Section \ref{sec:greeff}).

\subsubsection{Fermi KPT Rates}
\label{sec:fermi_rates}

We now derive the rates, $R_{ij}$ describing the transition from phase $i$ to phase $j$, that are needed in the KPT framework.
Via the master equation, which is mass conservation, we have that
\begin{equation} 
\frac{d \lambda_i}{dt} = \sum_j \lambda_j R_{ji} - \lambda_i R_{ij}  \, ,
\end{equation}
where $\lambda_i$ is the mass fraction of phase $i$.
As pointed out in Ref. \cite{GreefKPT}, it is generally preferable to describe the phase transition in terms of mass fractions because the volumes are also temperature and pressure dependent.

For the phase transition in Figure~\ref{fig:FKPT}, from an initial state with only one phase, this reduces to
\begin{equation} 
\frac{d \lambda_1}{dt} = - \lambda_1 R_{12}  \, .
\end{equation}
Performing the derivative on Equation~\eqref{eqn:FKPT} gives us
\begin{equation} 
\frac{d \lambda_1}{dt} = \frac{d \Delta G}{dt} \frac{d \lambda_1}{d \Delta G} 
= \frac{d \Delta G}{dt}  \frac{-\left( e^{4 \frac{\frac{\Delta G}{\Delta G_0} - 1}{B_0}}\right) \frac{4}{\Delta G_0 B_0}}{\left( e^{4 \frac{\frac{\Delta G}{\Delta G_0} - 1}{B_0}}+1\right)^2} 
= - \lambda_1 \frac{1}{b_{12}}\frac{d \Delta G}{dt} \frac{1}{e^{- \frac{\Delta G - w_{12}}{b_{12}}}+1}
\, ,
\end{equation}
where
\begin{align}
b_{12} & =  \frac{\Delta G_0 B_0}{4} \, , &
w_{12} & =  \Delta G_0 \, .
\end{align}
In passing, we note that in the special case where the change in Gibbs free energy of both phases is parallel in some regime, the rate of our model's phase transition would go to zero because of $\frac{d \Delta G}{dt} =0$.
In fact, some experimental data have shown phase transitions coming to a halt in a mixed state on the order of hours at a fixed pressure which is higher than the transition pressure predicted by the EOS.
Ref. \cite{Merkel:2020}, for example, conducted a quasi-static study of the $\alpha\to\epsilon\to\alpha$ phase transition in iron over a pressure range of 5--25 GPa and over the course of several hours found that on these time scales the two phases can co-exist over a range of pressures.
We will come back to this point below in Section \ref{sec:comparison}.

In principle, our present model could be generalized with an additional model parameter setting a minimum rate, i.e. by replacing $\frac{d \Delta G}{dt}$ with $\max\left(r_0,\frac{d \Delta G}{dt} \right)$.
This way, one could account for long time-scale physics not captured by the kinetic parameters $b$ and $w$.
That generalization, however, is beyond the present scope where our main focus is on dynamic conditions and time scales smaller than hours.

\subsection{Pressure rate dependence of the model parameters}

An indication that the energy barrier (parametrized here by $w$) and the phase transition speed (parametrized here by $b$) should in fact be strain (or pressure) rate dependent comes from the in-depth study of Ref. \cite{Blaschke:2025a}.
In that paper, a micro-structure dependent phase transition kinetics model was developed, where the transition rates depend on dislocation density, grain sizes, and other material parameters, as well as on strain rate.
The difficulty with the more physics based model of \cite{Blaschke:2025a} is that many of the required parameters are not known:
dislocation densities and grain sizes are not typically known for every sample and will also change under high rate deformation conditions.
The interfacial energy between grains of different phases is equally unknown in most cases \cite{Blaschke:2025a}.
We therefore attempt to reproduce the main features of \cite{Blaschke:2025a} within our present Fermi KPT model by adding two more parameters which capture the volumetric strain rate dependence of $w_{ij}$ and $b_{ij}$ in the following way:
\begin{align}
W_{ij} &= w_{ij}\left(1+\gamma_w^{ij}\dot\varepsilon\right)
\,,&
B_{ij} &= b_{ij}\left(1+\gamma_b^{ij}\dot\varepsilon\right)
\,, \label{eq:fermi_rate_dep}
\end{align}
where $\dot\varepsilon=\dot P/P$ is the pressure rate divided by current pressure and both new parameters, $\gamma_w^{ij}$ and $\gamma_b^{ij}$, therefore have the dimension of time.
In a first step we assume $\gamma_w^{ij}=\gamma_b^{ij}$ in order to see if one additional parameter is sufficient to fit the data.
The interpretation of $\gamma_w$ is also straightforward: it is the time required to get the phase transition started and as the strain rate increases, the phase transition therefore ``lags behind'' which is equivalent to an increase in the energy barrier.

The reason we assume the linear increase of $W_{ij}$ with normalized pressure rate is due to a comparison to experimental data on iron \cite{Smith:2013}, more specifically the transition from $\alpha$ to $\epsilon$ iron under ramp loading conditions.
Smith et al. in Ref. \cite{Smith:2013} studied the onset pressure (i.e. the pressure at which the PT starts) as a function of volumetric strain rate.
This effect is precisely captured by the increase of $W_{ij}$ in our model.
In Ref. \cite{Blaschke:2025a}, the micro-structure dependent model was successfully fit to that data, although many sets of micro-structure parameters could reproduce the same behavior.

\section{Comparative Assessment Methodology}
\label{sec:validate}

This section presents both experimental data obtained for the purpose of validating the Fermi KPT model, along with brief summaries of two models that can be used for comparison. These models, mentioned previously, are the Greeff KPT model \cite{Greeff:2016}, which is applicable to similar length and time scales as the Fermi KPT model, and the micro-structure (or micro) model \cite{Blaschke:2025a}, which includes more resolved physics about the nucleation and propagation of phase transitions. 

\subsection{The tin \texorpdfstring{$\beta - \gamma$}{beta - gamma} transition under shock loading}
\label{sec:tingundata}

To provide a calibration dataset for the $\beta-\gamma$ transition in tin, a series of gas-gun flyer-plate impact experiments were performed.
In short, a sabot is used to launch a disc of material (i.e., the flyer-plate) into a similarly shaped disc of the sample material (i.e., the target material) at high velocity, generating a planar shock wave in the sample.
The stress behind the shock front is controlled by both the flyer-plate material and the impact velocity.

The shock impact testing was carried out on a 19 mm bore light-gas gun.
Argon gas was pressurized inside of a wrap-around breech, and a dual o-ring projectile made of Lexan and fitted with a flyer plate was launched down the barrel.
The back of the flyer plate was fitted with syntactic foam to support the flyer during launch.
Projectile velocity immediately before impact was measured with a pair of laser light-gates spaced at a known distance.
The sample was secured on the front face of a catch-tank and aligned to the barrel axis before the shot using a laser.
With this method, the deviation from parallel at impact is typically submilliradian.

\begin{figure}[!h!t]
    \centering
    \includegraphics[width=0.65\linewidth]{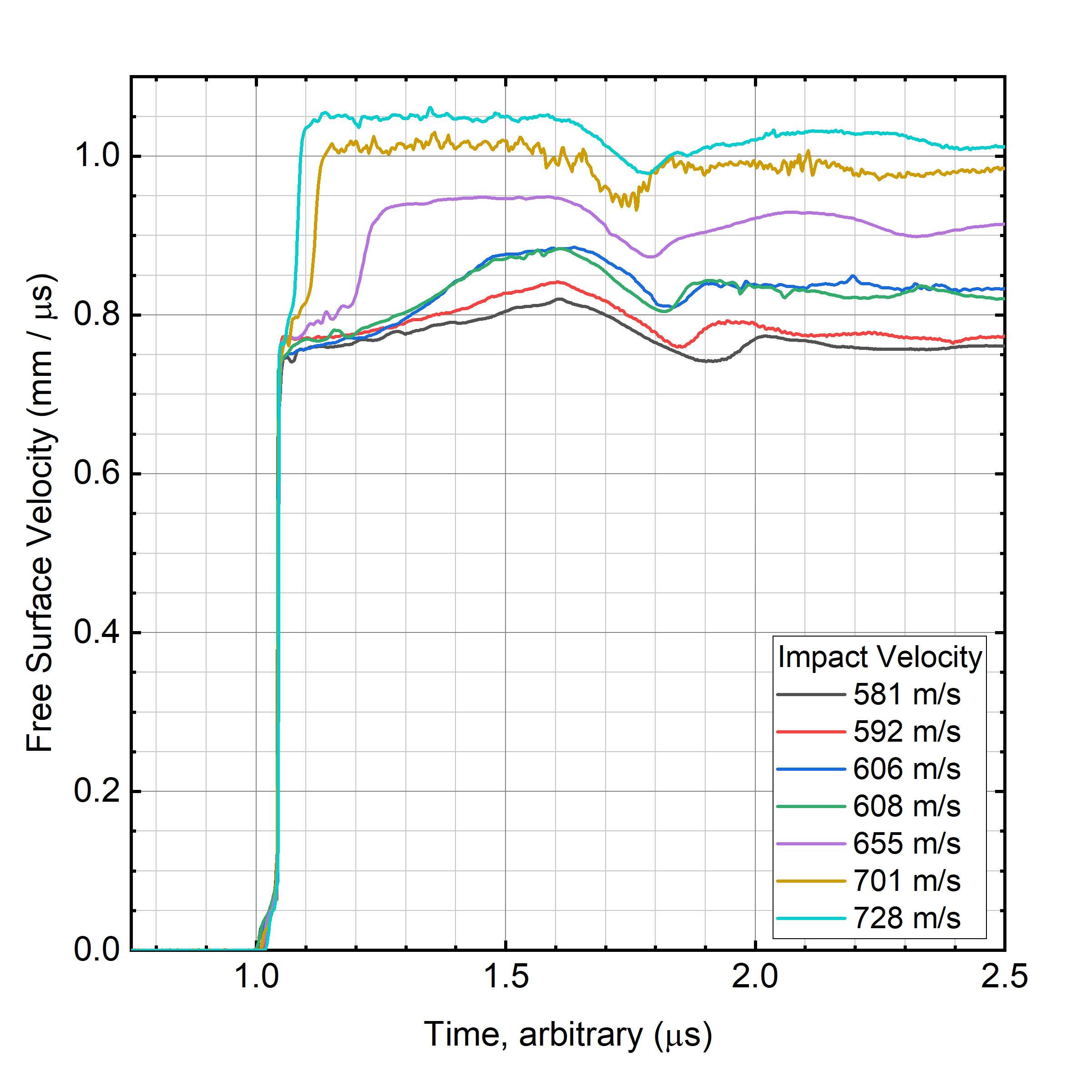}
    \caption{Experimental data for the tin flyer-plate impact experiments.  Profiles have been shifted in time to align at first shock breakout.}
    \label{fig:tingundata}
\end{figure}

In this series, the target samples were high-purity tin, of which the microstructure and high-rate strength properties have been previously described \cite{Nguyen:2024}.
The bulk tin material was machined into discs, 15.2~mm diameter by 3.1~mm thick.
For each experiment, the target sample was placed into a kinematic mount, and a laser was used to ensure the samples were concentric and normal to the gas-gun barrel.
This typically ensures a sub-milliradian tilt between the flyer-plate and target sample at impact.
For all experiments, the flyer-plates were tantalum, machined into discs 18.5~mm diameter by 1.5~mm thick.
These were mounted to the front of a lexan sabot.
A single fiber optic probe was placed behind the tin target at the center of the rear face to collect the free-surface velocity history via photon Doppler velocimetry (PDV).
The impact velocity was measured with a pair of laser light gates before the sample.
Impact velocities ranging between 581~m/s and 728~m/s were used to generate peak stresses from just over the $\beta-\gamma$ transition up to 12.5~GPa, to cover a range of transformation amounts and rates.
The free-surface velocity profiles were reduced from the raw data with conventional sliding Fourier transform techniques.
The experimental results are presented in Figure~\ref{fig:tingundata}.

All profiles exhibit the expected three-wave response for a shock-driven phase transformation under the overdrive limit - a small magnitude elastic precursor, a sharp main shock to the onset of the $\gamma$ phase transition, and then a third wave rising to the final peak shock state. 
The time between the main shock and the third wave is indicative of the time it takes for the phase transition to occur.
There is a clear pressure-dependency on the kinetics of the $\beta - \gamma$ transition, where the rate-of-rise of the third wave is proportional to the peak shock stress.
At stresses just above the transition onset, such as the 581~m/s and 592~m/s experiments, the transition is sluggish enough that a steady shock state is not reached in the duration of the loading pulse.
However, in the high stress experiments, such as the 728~m/s case, the transition completes in approximately 50~ns, almost being overdriven into a single shock profile.

\subsection{Review of the Greeff KPT model and its correlated model parameters}
\label{sec:greeff}

\begin{figure}[ht]
\centering
\includegraphics[width=0.5\linewidth]{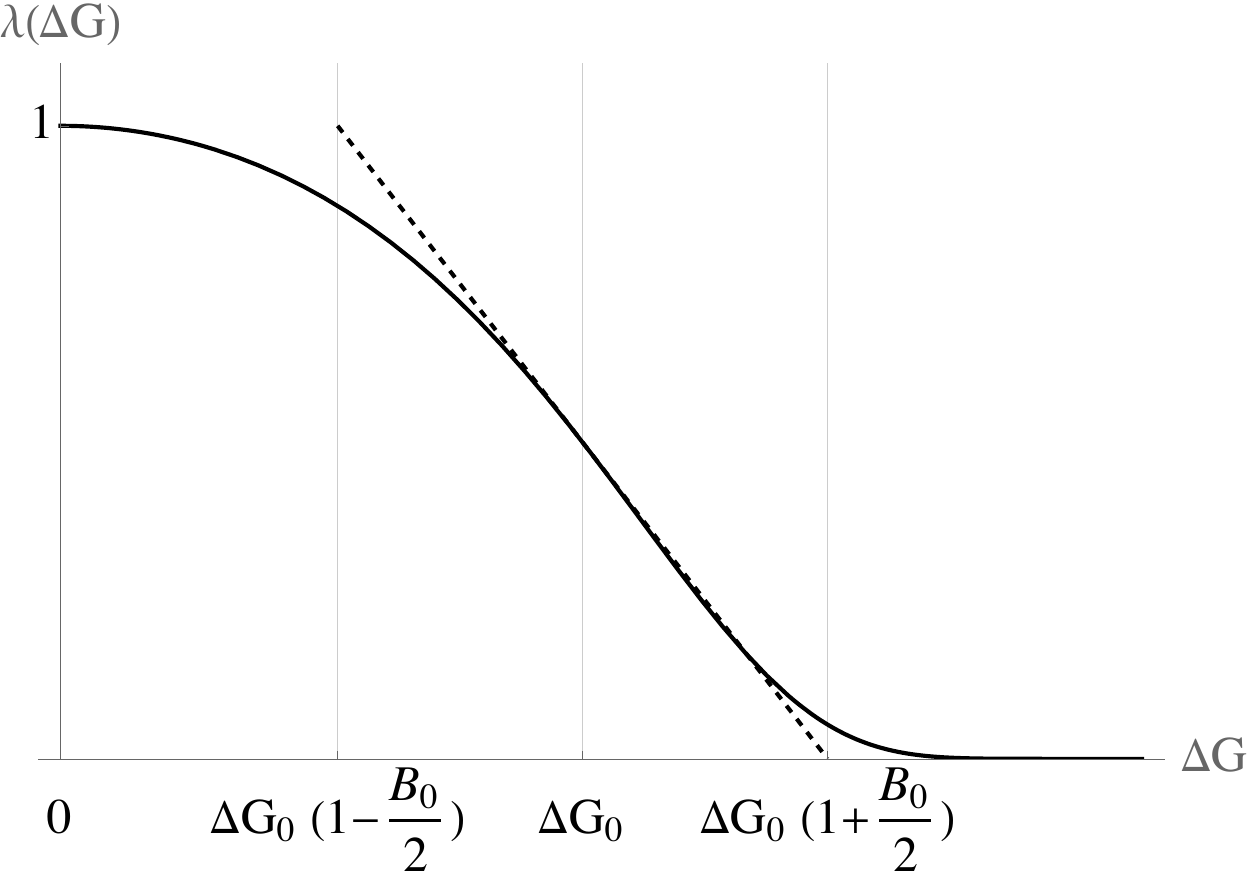}
\caption{\label{fig:GKPT}The Greeff KPT model. $\Delta G_0$ is related to the nucleation energy barrier and sets the energy scale of the model. The width of the phase transition is $B_0$ in terms of this energy scale. Note that $\Delta G$ is positive, a negative $\Delta G$ will not trigger a phase transition. The relations of $\Delta G_0$ and $B_0$ to  $C_1$ and $C_2$ in Equation~\ref{eqn:Greef} are given in the text.}
\end{figure}

Assuming a linearly growing energy difference in the Greeff model we have for the mass fraction of the initial phase (see References~\cite{Greeff:2016,GreefKPT}):
\begin{equation}
    \lambda_m(\Delta G) = \exp{\left(-\frac{C_1}{2} \left( \exp((\frac{\Delta G}{C_2})^2)-1 \right) \right)} \qquad \ \ \Delta G > 0 \, .
    \label{eqn:Greef}
\end{equation}
This curve is shown in Figure~\ref{fig:GKPT} with width, $B_0$, and halfway point, $\Delta G_0$, defined as before in Section \ref{sec:fermimodel}.
The parameters $\Delta G_0$ and $B_0$ can be related to the $C_1$ and $C_2$ parameters in Equation~\eqref{eqn:Greef} by
\begin{align}
B_0 & = \frac{2}{(C_1 + \ln{4}) \ln(\frac{C_1 + \ln{4}}{C_1})} \ ,\\
\Delta G_0 & = C_2 \sqrt{\ln(\frac{2 \ln{2}}{C_1} + 1)} \, .
\end{align}
We immediately notice that changing $C_1$ will not only change the slope, and thereby the width, $B_0$, of the phase transition, but it will also affect the halfway point $\Delta G_0$.
This coupling is undesirable but it is not the main problem.
The fact that this form restricts the width to the range
\begin{equation}
    0 < B_0 < \frac{2}{\ln{4}} \approx 1.4427 \, ,
\end{equation}
is, however, a major disadvantage, particularly in relatively slow transitions where the width of the transition would be large.
Additionally, the double-exponential in Equation \eqref{eqn:Greef} makes the mass fraction evolution of the Greeff KPT model very sensitive, not only to model parameter $C_2$, but also to small changes in the difference in Gibbs free energy, $\Delta G$, as a function of pressure.
We should also note that in the Greeff KPT model, $C_1=\nu_G C_2/\Delta\dot{G}$, where $\nu_G$ (not $C_1$) and $C_2$ are the two model parameters \cite{GreefKPT} and $\Delta\dot{G}$ is the time-derivative of the difference in Gibbs free energy between the phases as a function of pressure.
This dependence of $B_0$ and $\Delta G_0$ on the Gibbs free energy difference leads to an asymmetry in the hysteresis curve describing the pressure driven forward and reverse phase transition as we will see in Section \ref{sec:comparison} below.

Model parameter $\nu_G$ constitutes a scaling factor for the rate of the phase transition as can be seen from the following equation (shown here for an initial state with only one phase):
\begin{align}
\frac{d \lambda_1}{dt} &= - \lambda_1 R_{12}
\,,&
R_{12} &= -\nu_G\frac{\Delta{G}}{C_2}\exp{\left((\frac{\Delta G}{C_2})^2\right)}
\,,&& \Delta G>0
\,,
\end{align}
and $\lambda_1$ given above in \eqref{eqn:Greef}.
Even if $\Delta G$ were constant (and $\Delta\dot{G}\to0$), the rate of the phase transition would not vanish in the Greeff model.
Hence, a minimum rate, like we proposed earlier in Section \ref{sec:fermi_rates} for the Fermi model, is not necessary in the Greeff model whose rate is proportional to its model parameter $\nu_G$.

\subsection{Review of the micro-structure dependent model}
\label{sec:modelreview}
\subsubsection{Determining the volume fraction of the new phase}

In this subsection, we model the kinetic transformation by calculating the volume fraction of the new phase as a function of time, following (and summarizing) Ref. \cite{Blaschke:2025a}.
Table \ref{tab:modelparams_micro} summarizes the values we use in this paper for the model parameters whose meanings are detailed below.
For simplicity, we restrict ourselves here to ramp loading so that the current pressure is linearly increasing with time.
We account for homogeneous nucleation, nucleation on dislocations, and nucleation on grain boundaries, edges, and vertices.
Key quantities we required for all three nucleation mechanisms include the pressure derivative of the difference in Gibbs free energy across the phase boundary on the coexistence curve normalized by material density, $\Delta G'_P$ (which is dimensionless), and the interface speed of the phase transition, $c$.
The difference in Gibbs free energy is most conveniently calculated from an analytic equation of state, see Section \ref{sec:eos} below.

\begin{table}[ht]
\centering
\begin{tabular}{l|c||c}
 & $\beta$-tin / $\gamma$-tin & $\alpha$-iron / $\epsilon$-iron \\\hline
$\kappa$ [m$^3$/Js] & 1 & 1\\
$\beta$ [J/m] & 1e-10 & 1e-10\\
$\Delta P$ [GPa] & 10 & 10\\
$\gamma_{AM}$ [mJ/m$^2$] & 50 & 50\\
$\gamma_{AA}$ [mJ/m$^2$] & 70 & 70\\
$b_B$ [\r{A}] & 3.2 & 2.5\\
$\mu$ [GPa] & 18.4 & 81.6 \\
$\nu$ & 0.357 & 0.293\\
$D$ [cm] & 0.01 & 0.01\\
$\delta$ [cm] & 1e-8 & 1e-8\\
$\varrho_\text{dis}$ [m$^{-2}$] & 1e10 & 1e10
\\\hline
\end{tabular}
\caption{Summary of model parameters used here.}
\label{tab:modelparams_micro}
\end{table}

The interface speed is determined using a phase field model based on the Gibbs free energy of both phases.
In particular, by solving a time-dependent Ginzburg-Landau equation of the form
\begin{equation}
\frac{\partial \eta}{\partial t} = - \kappa \left(\frac{\partial G}{\partial \eta} - 2 \beta \frac{\partial^2 \eta}{\partial x^2} \right)\, ,
\end{equation}
where $\eta$ is the phase field variable one gets \cite{Blaschke:2025a}
\begin{align}
c &= 2\kappa\sqrt{3\beta\Delta G'_P\Delta P}\,x
\,, \nonumber\\
x &\approx \frac{P-P_\text{trans}}{\Delta P}
\,. \label{eq:interfacespeed}
\end{align}
$P_\text{trans}$ is the pressure on the coexistence curve of the $\beta$ and $\gamma$ phases , $\Delta P$ is the pressure difference between the coexistence curve and the $\beta\to\gamma$ spinodal, and $\kappa$ and $\beta$ are the kinetic and gradient energy coefficients within the time dependent Ginzburg-Landau equation with dimensions of volume/energy-time and energy/length, respectively.
Since we lack accurate values for these coefficients, we must regard the prefactor of $c$ as a tuneable model parameter.
In particular,
\begin{align}
c &= c_\text{pref} \left(P-P_\text{trans}\right)
\,,&
c_\text{pref} &= 2\kappa\sqrt{\frac{3\beta\Delta G'_P}{\Delta P}}
\,.
\end{align}

The basic concept of the Kolmogorov-Johnson-Mehl-Avrami (KJMA) kinetic theory \cite{Kolmogorov:1937,Johnson:1939,Avrami:1939, Avrami:1940,Avrami:1941} is the extended volume fraction, $\lambda_E$, which is the sum of the volume of all growing nuclei without accounting for nucleus-nucleus impingement.
If the nucleation sites are randomly distributed, then the physical volume fraction can be approximated as
\begin{equation}
\lambda_V = 1-\exp{(-\lambda_E)}
\,.
\end{equation}
In the following, we assume ramp loading (i.e. a constant pressure rate $ \dot{P}$ driving the phase transformation) to simplify the integrations in the nucleation equations below.

\subsubsection{Homogeneous nucleation}

The extended volume fraction due to homogeneous nucleation is computed from \cite{Blaschke:2025a}
\begin{align}
\lambda_E^\text{hom} &= \int_0^t dt' \left(t^2-t'^2\right)^3 \dot{P} ^3\exp\left(a-\frac{b}{(\dot{P} t')^2}\right)
\,,\nonumber\\
a &= \log\left[\frac{4\pi \nu_D n}{3}\left(\frac{c_\text{pref} }{2}\right)^3\right]
\,, &
b &=  \left(\frac{16\pi}3\right)\frac{\gamma_{AM}^3}{\left(\Delta G_{P}'\right)^2k_\text{B}T}
\end{align}
where $\nu_D\sim10^{13}$s$^{-1}$ (Debye frequency) and $n$ is the number density of atoms, which can be estimated as $N_A/\bar V$, i.e. Avogadro's number divided by the volume per mol averaged over both phases.
The average interfacial energy between grains of $\beta$ and $\gamma$ tin, another input parameter, is denoted by $\gamma_{AM}$.

\subsubsection{Heterogeneous nucleation via dislocations and grain boundaries}

The extended volume fraction due to nucleation on dislocations was derived in Ref. \cite{Blaschke:2025a} which was based in part on \cite{Cahn:1957}. Summarizing the governing equations, we have
\begin{align}
\lambda_E^\text{dis} &= \int_0^t dt'  \left(t^2-t'^2\right)^3 \dot{P}^3 \exp\left(a-\frac{b}{(\dot{P} t')^2}\right)
\,,\nonumber\\
a &= \log\left[\varrho_\text{dis}b_B^2\frac{4\pi \nu_D n}{3}\left(\frac{c_\text{pref}}{2}\right)^3\right]
\,,\nonumber\\
b &= f_\text{dis}(\alpha(\dot{P}, t')) \left(\frac{16\pi}3\right)\frac{\gamma_{AM}^3}{\left(\Delta G_{300}'\right)^2k_\text{B}T}
\,,\nonumber\\
f_\text{dis}(\alpha) &\approx (1-\alpha)\left(1-\frac45\alpha\right)\theta(1-\alpha)
\,,\nonumber\\
\alpha (\dot{P},t') &= \frac{\mu b^2 \tilde\kappa \Delta G}{2\pi^2\gamma^2_\text{AM}}\approx  \frac{\mu b^2  \Delta G'_{300}\dot{P} t'}{2\pi^2\gamma^2_\text{AM}}\frac{(1-\nu/2)}{(1-\nu)}
\,,
\end{align}
with (average) shear modulus $\mu=18.4$ GPa for $\beta$-tin and its Poisson ratio $\nu=0.357$ and $\tilde\kappa\approx(1-\nu/2)/(1-\nu)$ (average over edge and screw dislocations in the isotropic limit).
$b_B$ denotes the Burgers vector length.


The volume fraction evolution due to nucleation on grain boundaries, edges (triple points), and corners (4-grain junctions) was derived in Ref. \cite{Blaschke:2025a} based in part on earlier work of Ref. \cite{Clemm:1955}.
Ref. \cite{Blaschke:2025a} showed that for ramp loading in iron the leading contribution to nucleation from grains is given by the grain boundaries. 
Thus, in the comparison here we neglect nucleation on grain edges and corners (whose contribution is even smaller than that from edges).
The equations describing nucleation from grain boundaries summarize as:
\begin{align}
\lambda_{GB,E}^\text{grain}(t) &= 2s_{GB}\frac{r(t,0)}{D}\left[1-\int_0^1dx\exp\left\{-\pi r^2(t,0)\int_0^{t(1-x)} dt' I_{GB}^\text{grain}(t')\left[\left(1-\frac{t'}{t}\right)^2-x^2\right]\right\}\right]
\,,\nonumber\\
I_{GB}^\text{grain}(t) &\approx \frac{\nu_D N_A}{\bar{V}} \delta^{3-d} \exp\left(-\left(\frac{16\pi}3\right)\frac{\gamma_{AM}^3}{\left(\Delta G_{P}'\right)^2k_\text{B}T}\frac{f_{GB}^\text{grain}(k)}{\dot{P}^2t^2}\right)
\,,\nonumber\\
f_{GB}^\text{grain}(k) &=\max\left[0,\frac12\left(2 - 3 k + k^3\right)\right]
\,,
\end{align}
where $s_{GB}=0.75\left(1+2\sqrt{3}\right)\approx3.3481$
for bct crystals with $c/a<\sqrt{2}$ (where the Wigner-Seitz cells are truncated octahedra), which is appropriate for $\beta$-tin.
Furthermore, $k=0.5\gamma_{AA}/\gamma_{AM}=0.7$ if we assume $\gamma_{AA}=1.4\gamma_{AM}$ (as a rough estimate).

Using Mathematica, we can integrate over $t'$ in $\lambda_{GB,E}^\text{grain}$, and for the case of grain boundaries we obtain

\begin{align}
\lambda_{GB,E} &= 2\frac{s_{GB}}D r(t,0)\left[1-\int_0^1dx\exp\left(-\pi r(t,0)^2 \nu_DN_A \frac{\delta}{\bar{V}}\exp\left[-\pi r(t,0)^2\delta\frac{\nu_DN_A}{\bar{V}}J_2(t,x)\right]\right)\right]
\,,\nonumber\\
r(t,0)[\text{cm}] &= \left(\frac{c_\text{pref}}{2}\right){\dot{P}t^2}
\,,\nonumber\\
J_{GB}(t,x) &= \int_0^{t(1-x)} dt' \left[\left(1-\frac{t'}t\right)^2-x^2\right]e^{-A/t'^2} 
\nonumber\\
&= \frac{e^{-A/B^2}}{3t}(x-1)\left(2A+(x-1)(2x+1)t^2\right) + \frac{A}{t}E_1\left(\frac{A}{B^2}\right)
+\sqrt{A\pi}\left(2A+3(x^2-1)t^2\right)\frac{\text{erfc}\left(\frac{\sqrt{A}}{B}\right)}{3t^2}
\,,\nonumber\\
A[\mu\text{s}^2] &= \left(\frac{16\pi}3\right)\frac{\gamma_{AM}^3}{\left(\Delta G_{P}'\right)^2k_\text{B}T}\frac{f_2^\text{grain}(k) }{\dot{P}^2}
\,,\nonumber\\
B &= t(1-x)
\end{align}
where the exponential integral $E_1(z)=\int_z^\infty \frac{e^{-t}}{t} dt$  and the complementary error function erfc$(x)=1-\text{erf}(x)$.



The model reviewed above, which is available as an open source Python research code \cite{pyptkinetics}, has the full formulation including nucleation of grain edges and corners. As with all models, it has advantages and disadvantages:
On the one hand, we include many physics-based mechanisms which can be expected to give accurate predictions regarding phase transition kinetics, if (and only if) (1) the micro-structure prior to loading was measured and characterized sufficiently and (2) an accurate model for micro-structure evolution is included.
For recent developments on dislocation density evolution, see for example Ref. \cite{Hunter:2022}.
On the other hand, it requires numerical integration of various terms and therefore can never be as fast computationally as a simple phenomenological model.
Additionally, the present micro-structure dependent model can only reproduce a hysteresis curve at elevated pressure rate, but not in the quasi-static limit (in contrast to experimental data such as \cite{Giles:1971,Merkel:2020}) as highlighted below in Figure \ref{fig:volfracTin}.

\section{Results and Comparison using the Fermi KPT model}

\subsection{Analytic equations of state for iron and tin}
\label{sec:eos}

For application of the Fermi KPT model and the other models described in Section \ref{sec:validate}, an equation of state (EOS) is required. 
For ramp loading conditions, it is convenient to use analytic EOSs.
For ($\alpha$ and $\epsilon$) iron, we use the analytic EOS of Boettger and Wallace \cite{Boettger:1997}, as described in Ref. \cite{Blaschke:2025a} and implemented in the 1-dimensional research code, ``PyPTkinetics'', used for the ramp-loading calculations, which is available on GitHub \cite{pyptkinetics}.

\begin{table}[ht]
\centering
\begin{tabular}{l|l|l||l|l|l}
tin phase & $\beta$ & $\gamma$ & iron phase & $\alpha$ & $\epsilon$\\\hline
$\rho_0$ [kg/m$^3$] & 7285 & 7271 & $V_{300}$ [cm$^3$/mol] & 7.0093 & 6.73\\
$T_\text{ref}$ [K] & 298 & 298 & $T_\text{ref}$ [K] & 300 & 300 \\
$K_0$ [GPa] & 52.90 & 38.78 & $V^*$ [cm$^3$/mol] & 7.0047 & 6.5984 \\
$\left(\diff{K}{P}\right)_0$ & 5.3345 & 6.0532 & $B^*$ [GPa] & 176.64 & 181.5 \\
$\alpha_0$ [$10^{-5}$/K] & 7.2977 & 10.85405 & $B_1^*$ & 4.7041 & 5.74 \\
$C_{V0}$ [J/kg/K] & 214.9 & 216.1 & $\Phi^*$ [J/mol] & 0 & 5533 \\
$E_0$ [J/kg] & 65800 & 102500 & $\theta_0^{300}$ [K] & 301 & 261 \\
$S_0$ [J/kg/K] & 441.9 & 505.1& $\theta_2^{300}$ [K] & 420 & 364 \\
& & & $\gamma_0^{300}$ & 1.82 & 2.8 \\
& & & $T_m$ & 1135 & 1135 \\\hline
$P_\text{trans}$ [GPa] & \multicolumn{2}{|c|}{9.4} & & \multicolumn{2}{c}{13}\\
$\Delta G'_P$(300K) & \multicolumn{2}{|c|}{0.0245} & & \multicolumn{2}{c}{0.0519}\\\hline
\end{tabular}
\caption{Parametrization of the EOS for $\beta$ and $\gamma$ tin as well as for $\alpha$ and $\epsilon$ iron.
For the functional form of the EOS of iron and the meaning of its parameters, see Refs. \cite{Boettger:1997,Blaschke:2025a}.
The last two rows show quantities derived from the EOS: the transition pressure and a dimensionless quantity needed for the micro-structure dependent model of Sec. \ref{sec:modelreview}.
For iron it was derived in Ref. \cite{Blaschke:2025a}, for tin we derive it below in Eq. \eqref{eq:DeltaGP}.}
\label{tab:tineosparam}
\end{table}

For tin we use an extended Vinet EOS \cite{Lemke:2016} in our research code for ramp loading conditions ``PyPTkinetics'', which was parametrized for tin in the publicly available Los Alamos report \cite{Mattsson:2023Vinetpara}  (see also \cite{Mattsson:2016}).
However, this tin EOS has not been previously presented in a peer-reviewed publication, so we summarize the model here.
In particular, the pressure-temperature relation can be written as:
\begin{align}
P(\rho,T) &= P_\text{ref}(\rho) + \alpha_0 K_0\left(T-T_\text{ref}\right)
\,,\nonumber\\
P_\text{ref}(X) &= \frac{3 K_0}{X^2}\left(1-X\right)\exp\left[\eta_0\left(1-X\right)\right] 
\,,\nonumber\\
X &= \left(\rho_0/\rho\right)^{1/3}
\,,\qquad \eta_0 =\frac32\left(\left(\diff{K}{P}\right)_0-1\right)
\,,
\end{align}
with parameters including the thermal expansion coefficient $\alpha_0$ and isothermal Bulk modulus $K_0$ at the reference temperature $T_\text{ref}$, and equilibrium density $\rho_0$ at that temperature (i.e. $P(\rho_0,T_\text{ref})=0$).
Assuming $\alpha_0$, $K_0$, and the constant-volume heat capacity $C_{V0}$ are constant, the entropy can be written as
\begin{align}
S(\rho,T) &= S_0 + \alpha_0 K_0\left(V-V_0\right) + C_{V0}\ln\frac{T}{T_\text{ref}}
\,,\nonumber\\
C_{V0} &= \left(T\diff{S}{T}\Bigg|_V\right)\left(\rho_0,T_\text{ref}\right)
\,,
\end{align}
where $V=1/\rho$ and $V_0=1/\rho_0$.
The internal energy is given by
\begin{align}
E(\rho,T) &= E_0 + 9\frac{K_0V_0}{\eta_0^2} \left(1-\exp\left[\eta_0(1-X)\right]\left[1-\eta_0(1-X)\right]\right)
- \alpha_0 K_0V_0\left(1-X^3\right)T_\text{ref} + C_{V0}\left(T-T_\text{ref}\right)
\,.
\end{align}
The Gibbs free energy follows from a Legendre transformation, i.e.
\begin{align}
G(P,T) &= E(\rho(P,T),T) - TS(\rho(P,T),T) + \frac{P}{\rho(P,T)}
\,.
\end{align}
Parameters of the reference state for both phases are given in Table \ref{tab:tineosparam}.
With these parameters, the transition pressure at room temperature is about $P_\text{trans}=9.4$ GPa, i.e. \[G^\gamma(P_\text{trans},300)-G^\beta(P_\text{trans},300)=0
\,.\]
We then proceed to calculate
\begin{align}
\Delta G_P'(P_\text{trans},300) &:= \frac12\left(\rho^\beta(P_\text{trans},300)+\rho^\gamma(P_\text{trans},300)\right)\diff{\left(G^\beta(P,300)-G^\gamma(P,300)\right)}{P}\Bigg|_{P_\text{trans}}
\nonumber\\
&\approx 0.0244968
\,,\label{eq:DeltaGP}
\end{align}
a dimensionless quantity we need in our computations of the $\gamma$-tin volume fraction, see Section \ref{sec:modelreview}.

\subsection{Simulating ramp loading in iron and tin}
\label{sec:comparison}

\begin{figure}[!h!t]
\centering
\includegraphics[width=0.5\textwidth]{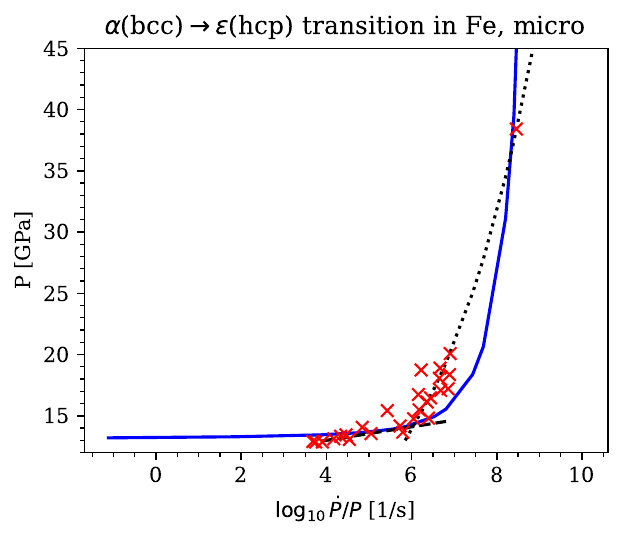}%
\includegraphics[width=0.5\textwidth]{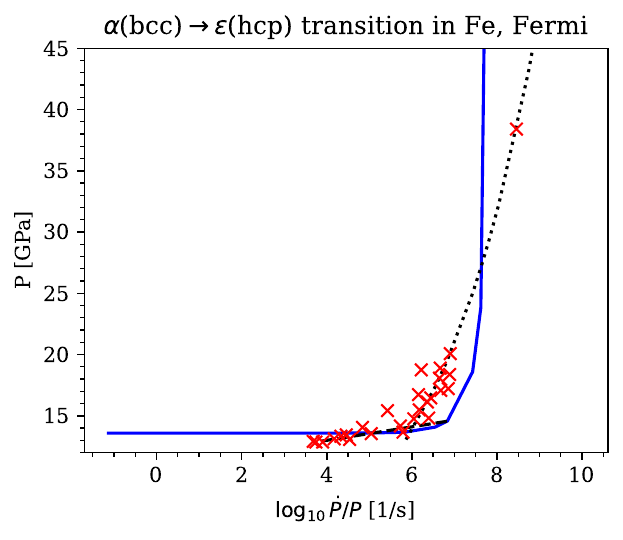}
\put(-0.99\columnwidth,0.42\columnwidth){\textbf{(a)}}
\put(-0.49\columnwidth,0.42\columnwidth){\textbf{(b)}}

\includegraphics[width=0.5\textwidth]{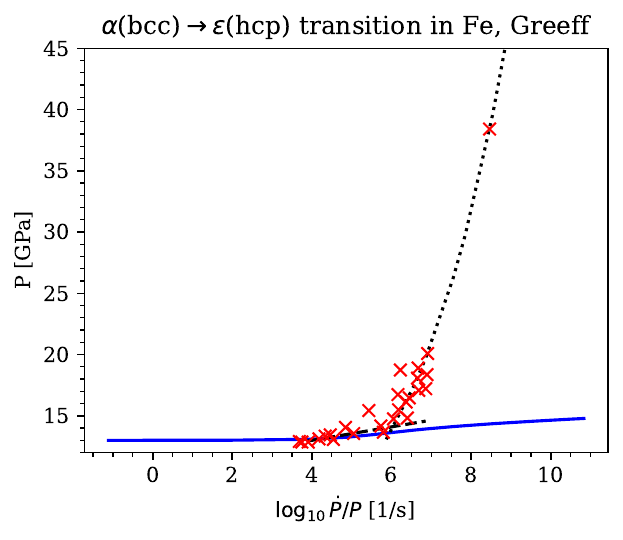}%
\includegraphics[width=0.5\textwidth]{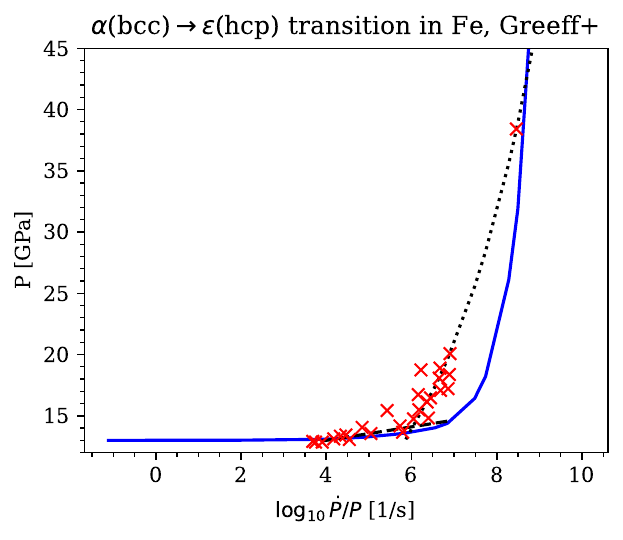}
\put(-0.99\columnwidth,0.42\columnwidth){\textbf{(c)}}
\put(-0.49\columnwidth,0.42\columnwidth){\textbf{(d)}}
\caption{The generalized rate dependent Fermi model with additional parameters $\gamma_w^{ij}$ and $\gamma_b^{ij}$ is able to capture the rate dependence of the onset pressure of the $\alpha\to\epsilon$ iron phase transition under ramp loading (see (b) on the top right in comparison with the full micro-structure dependent model (a) on the top left).
In particular, we compare our simulations to experimental results of \cite[fig. 6b]{Smith:2013}.
The dashed (dotted lines) represent linear (non-linear) fits to the latter (red crosses), where the non-linear fit is valid for strain rates above 10$^6$/s and the linear fit is valid between $\sim5\!\times\!10^3$--10$^6$/s.
The generalized Fermi model shown here in blue on the top right (b) performs almost as well as the micro-structure dependent model of Ref. \cite{Blaschke:2025a} (a).
In the bottom row, we show how the Greeff model performs in comparison, noting that its parameters need to be made rate dependent in order to capture the non-linear regime (bottom left without rate dependence, (c), bottom right with rate dependence, (d)).}
\label{fig:compareSmith}
\end{figure}

\begin{figure}[!h!t]
\centering
\includegraphics[width=0.5\textwidth,trim=0 0 4.cm 0, clip]{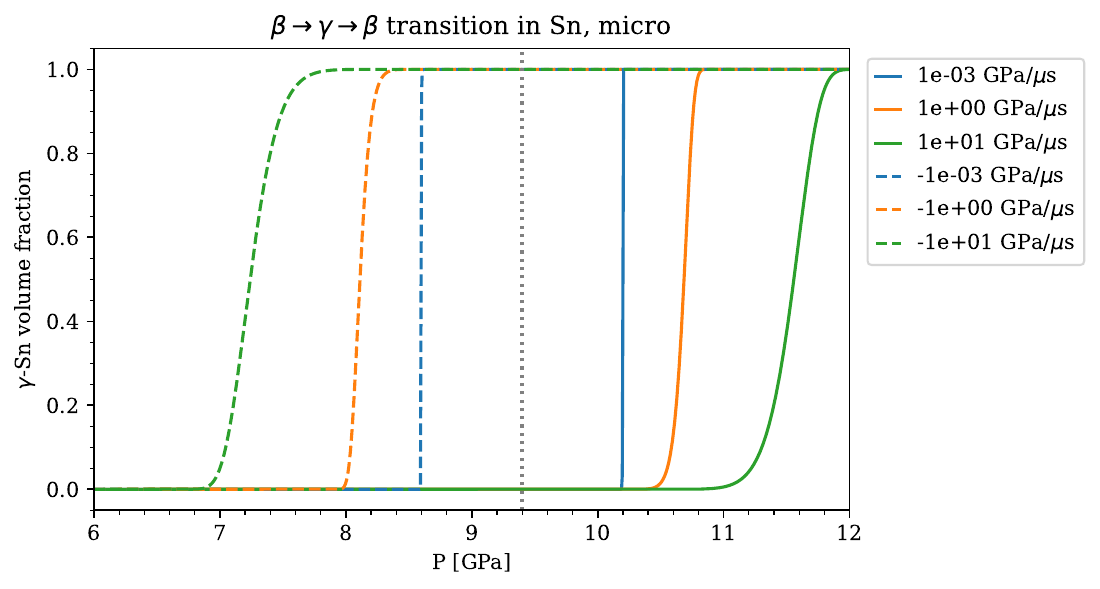}%
\includegraphics[width=0.5\textwidth,trim=0 0 4.cm 0, clip]{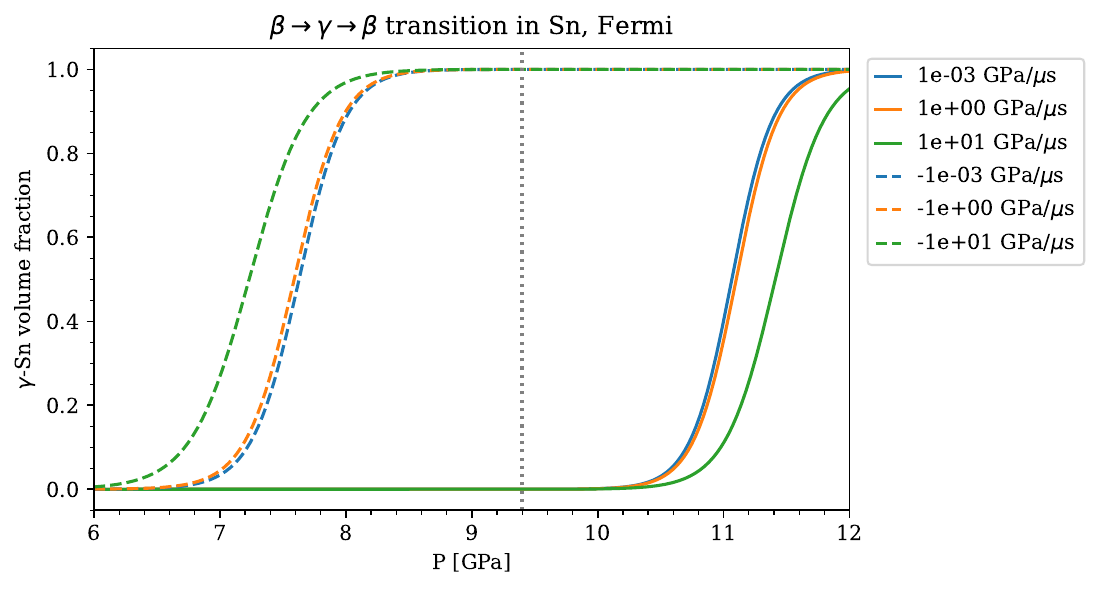}
\put(-0.99\columnwidth,0.32\columnwidth){\textbf{(a)}}
\put(-0.49\columnwidth,0.32\columnwidth){\textbf{(b)}}

\includegraphics[width=0.65\textwidth]{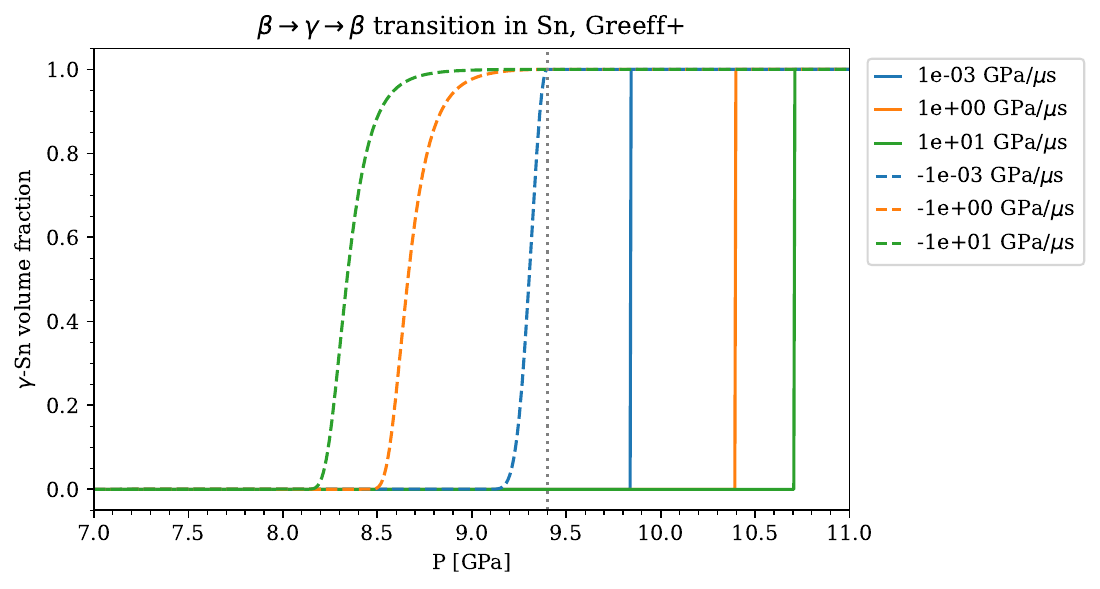}
\put(-0.66\columnwidth,0.32\columnwidth){\textbf{(c)}}
\caption{While the micro-structure dependent model (a) can only reproduce a hysteresis curve at elevated pressure rate, the generalized Fermi model (b) with additional parameters $\gamma_w^{ij}$ and $\gamma_b^{ij}$ yields the expected increase in hysteresis size for the forward ($\beta\to\gamma$ tin, solid lines) and reverse ($\gamma\to\beta$ tin, dashed lines) PT under ramp loading conditions of increasing pressure rate.
The Greeff model (c), even with rate dependent parameters, exhibits an unphysical asymmetry in hysteresis curve.
The vertical dashed gray line indicates the transition pressure of the analytic EOS presented in Section \ref{sec:eos}.}
\label{fig:volfracTin}
\end{figure}

All three models, i.e., the Greeff model summarized in Section \ref{sec:greeff},  the micro-structure dependent model summarized in Section \ref{sec:modelreview}, and the generalized Fermi model presented in Section \ref{sec:fermimodel}, have been implemented in the Python research code \cite{pyptkinetics} (originally developed for \cite{Blaschke:2025a}) by one of us (DNB).
This research code uses the analytic equations of state for iron and tin outlined above in Section \ref{sec:eos}.

In Ref. \cite{Blaschke:2025a}, some of us have shown a comparison between the micro-structure dependent model and ramp loading data of the $\alpha\to\epsilon$ iron transition of Smith et al. \cite{Smith:2013}.
In Figure \ref{fig:compareSmith} we compare all three models with that data set.
Note that the Greeff model does not have any rate dependence and therefore cannot capture the non-linear regime, see Figure \ref{fig:compareSmith}(c).
However, it is straightforward to make the two model parameters rate dependent by introducing two additional parameters in the same way as in the Fermi model --- see Equation \eqref{eq:fermi_rate_dep}, upon which a better match with the Smith data is achieved, see Figure \ref{fig:compareSmith}(d).

In Figure \ref{fig:volfracTin}, we show the results of a ramp-loading simulation for tin using the same research code.
In this case, we used the extended Vinet EOS parametrized by one of us (AEM) \cite{Mattsson:2023Vinetpara}, as summarized above in Section \ref{sec:eos}.
The figure shows the volume fraction of the $\gamma$ phase as a function of pressure, which under ramp loading increases (decreases) with time during the $\beta$ to $\gamma$ (resp. $\gamma$ to $\beta$) phase transition.
Note that in contrast to the micro-structure dependent model of Ref. \cite{Blaschke:2025a}, our present Fermi model can reproduce a pronounced hysteresis curve between the forward and reverse transformation via non-zero $w_{ij}$ and $b_{ij}$ even in the quasi-static limit (close to zero strain rate), an effect that is seen in experimental data (see e.g. \cite{Giles:1971,Merkel:2020}):
Even under quasi-static conditions the two phases can coexist over a range of pressures (non-zero $b_{ij}$) and the starting pressure for the forward transformation differs from the starting pressure for the reverse transformation (non-zero $w_{ij}$).
The Greeff model (even with its rate-dependent generalization), on the other hand, shows an unphysical asymmetric hysteresis curve (see Section \ref{sec:greeff}), where at low strain rates partial transitions are captured only for the backward transition but not for the forward transition.
For this reason, its rate-dependent generalization is not considered in the hydrocode simulations of the next section as the Fermi KPT model already addresses all shortcomings of the Greeff model.

To summarize, we expect that the newly derived rate dependent Fermi KPT model performs best across all pressure rate scales and proceed to calibrating this model further for tin in order to compare with the new experimental data presented above in Section \ref{sec:tingundata}.

\subsection{Simulating shock data on tin}

\begin{figure}[!ht]
\centering
\includegraphics[width=0.72\textwidth,trim=0 0 0 20.pt, clip]{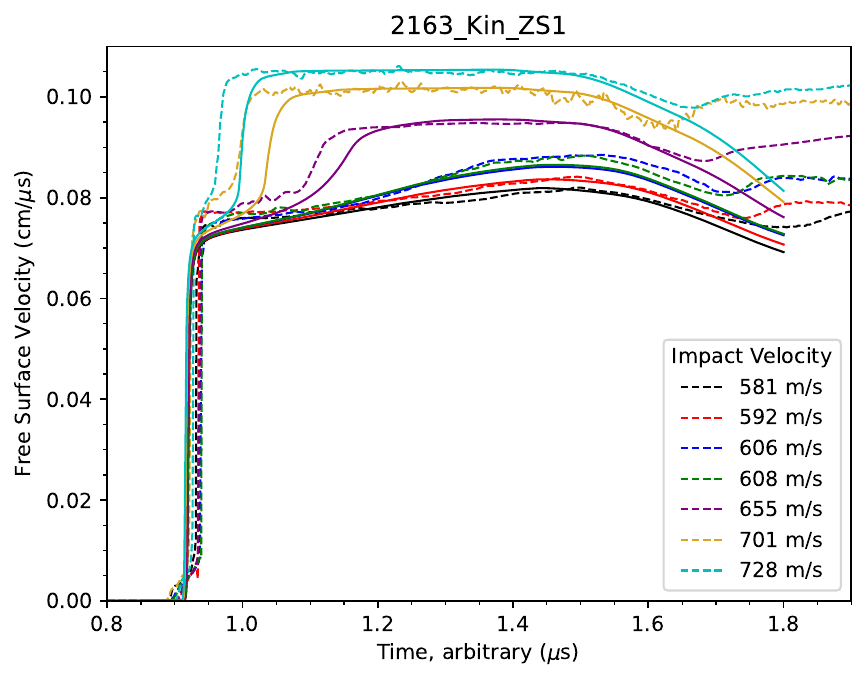}
\caption{We compare shock data on tin (dashed lines) with FLAG simulations (solid lines) using the Greeff model.
In order to overlay the simulation data with the experimental curves, time has been shifted (indicated by `arbitrary' in the axis label), i.e. $t=0$ has no special meaning.
The model parameters leading to this best fit were: $C_2=2$e3 J/kg, $C_1=\left(0.1\,\mu\text{s}^{-1}\right)*C_2/\Delta\dot{G}$,
as well as PTW flow stress scale factors 0.65 for $\beta$-tin and 0.1 for $\gamma$-tin, as calibrated in Ref. \cite{Prime:2026a}.
}
\label{fig:fitgreeff}
\end{figure}

\begin{figure}[!ht]
\centering
\includegraphics[width=0.72\textwidth,trim=0 0 0 20.pt, clip]{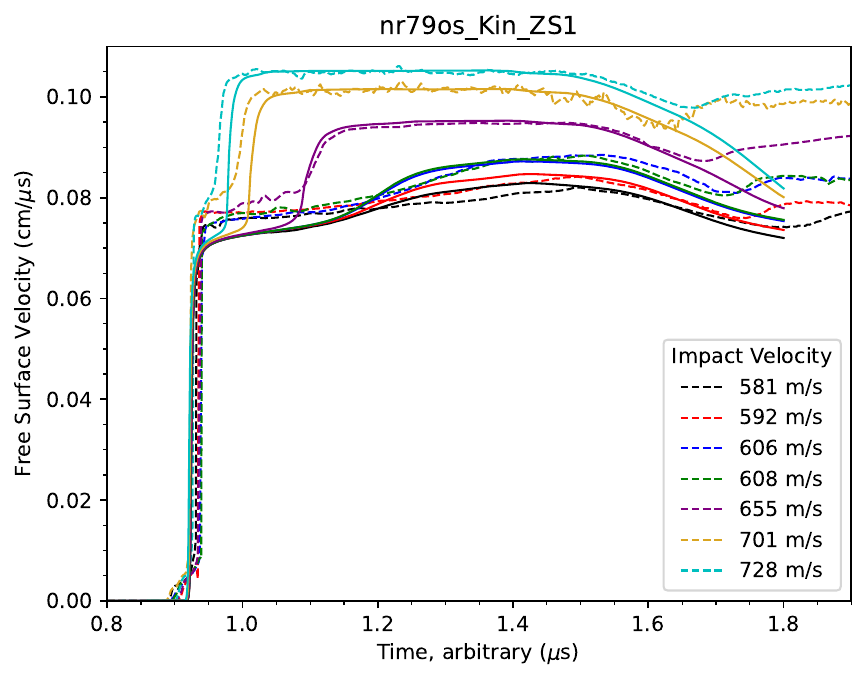}
\caption{We compare shock data on tin (dashed lines) with FLAG simulations (solid lines) using the Fermi model with $w=4$e3 J/kg, $b=3$e2 J/kg, $\gamma_b=\gamma_w=0$ (no rate dependence),
as well as PTW flow stress scale factors 0.65 for $\beta$-tin and 0.1 for $\gamma$-tin (i.e. the same scale factors as in Fig. \ref{fig:fitgreeff}).}
\label{fig:fitFermi}
\end{figure}

\begin{figure}[!ht]
\centering
\includegraphics[width=0.72\textwidth,trim=0 0 0 20.pt, clip]{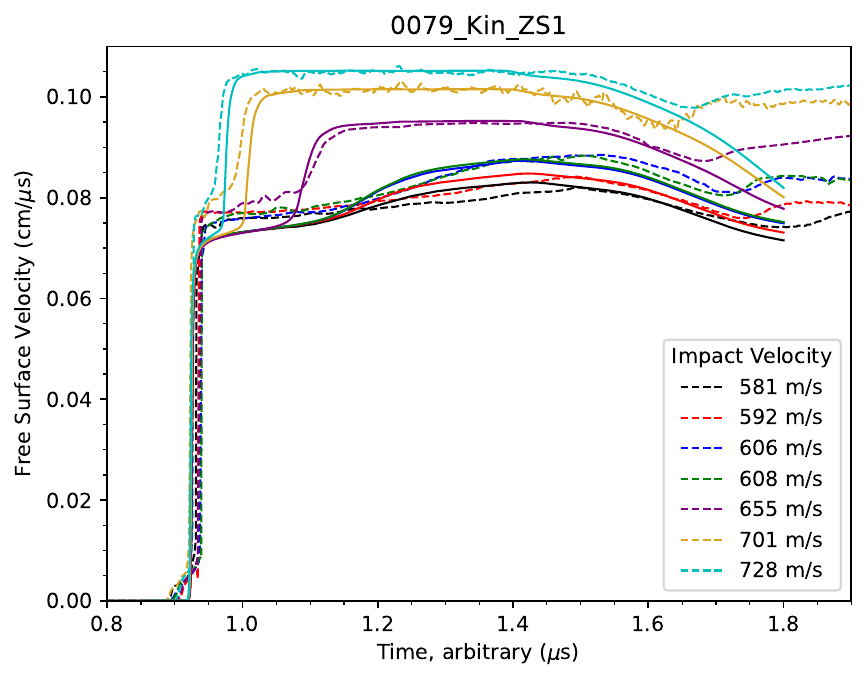}
\caption{We compare shock data on tin (dashed lines) with FLAG simulations (solid lines) using the rate dependent Fermi model with $w=4$e3 J/kg, $b=3$e2 J/kg, $\gamma_w=9$ $\mu$s, and $\gamma_b=9$ $\mu$s.
The fit was further improved by tweaking the PTW strength model, i.e. the flow stress scale factor of the $\gamma$ phase was increased to 0.2, while the scale factor for the $\beta$ phase was reduced to 0.53.
Those numbers are the upper/lower bounds according to Ref. \cite{Prime:2026a}.}
\label{fig:fitFermimod2}
\end{figure}

\begin{figure}[!ht]
\centering
\includegraphics[width=0.5\textwidth,trim=0 0 0 19.pt, clip]{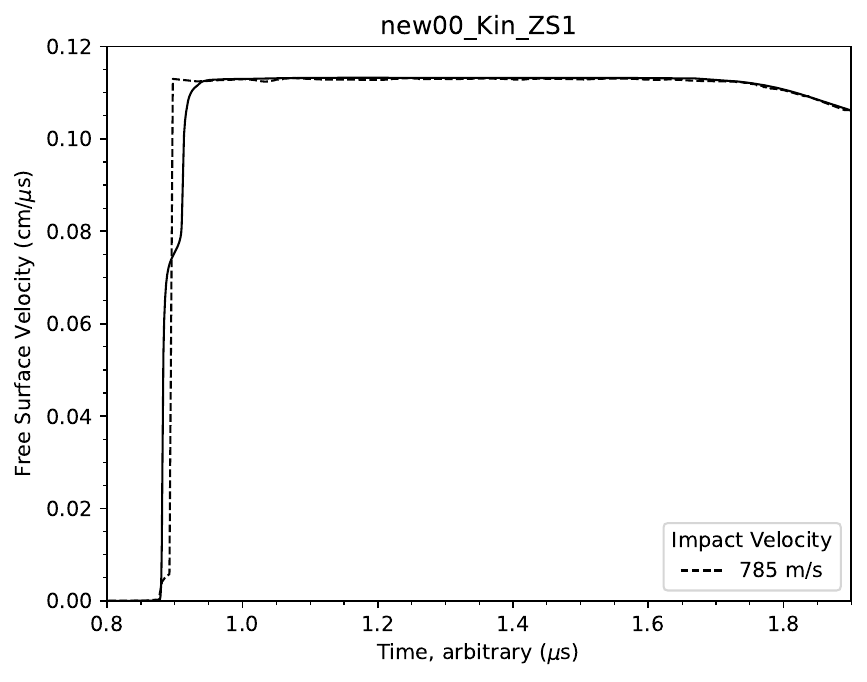}%
\includegraphics[width=0.5\textwidth,trim=0 0 0 19.pt, clip]{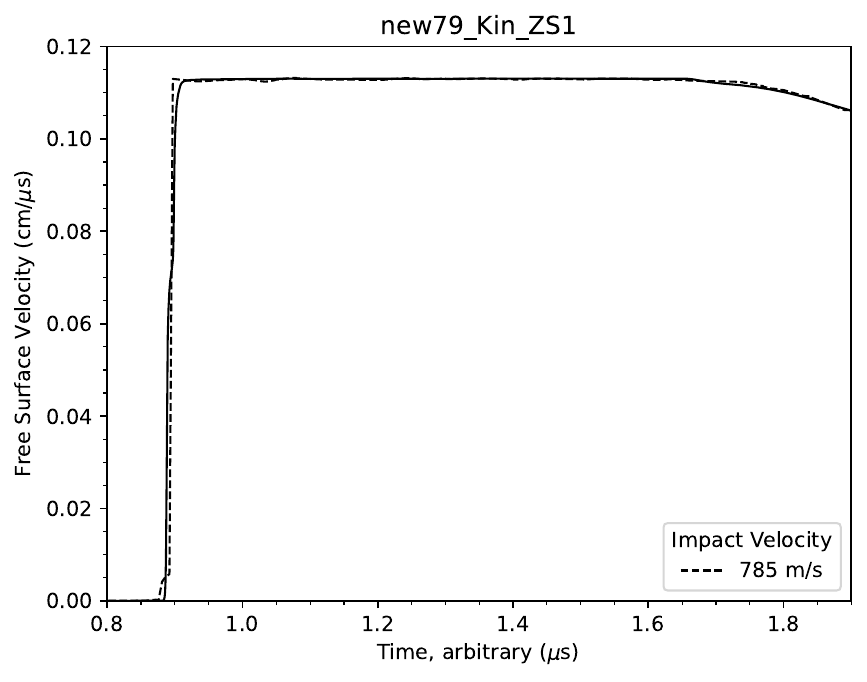}
\put(-0.99\columnwidth,0.01\columnwidth){\textbf{(a)}}
\put(-0.49\columnwidth,0.01\columnwidth){\textbf{(b)}}
\caption{In this additional comparison for the limiting case of an overdriven phase transition, we compare shock data on tin (dashed lines) from an even higher impact velocity with FLAG simulations (solid lines).
The experimental data (dashed line) shows a vertical single shock rise with no evidence of the transition due to the high impact velocity.
Nonetheless, the Greeff model (a) shown on the left (solid line) exhibits a pronounced kink and is thus unable to adequately capture this case.
With the rate dependent Fermi model (b) with $w=4$e3 J/kg, $b=3$e2 J/kg, $\gamma_w=9$ $\mu$s, and $\gamma_b=9$ $\mu$s shown on the right (solid line), the kink is much smaller and the agreement with the experimental data much better.
In passing we note that the changed scale factors (0.53 for $\beta$ and 0.2 for $\gamma$ tin) slightly improve the fit for the Fermi model also in this case.
}
\label{fig:fitFermimod2b}
\end{figure}

We performed 1-dimensional plate impact simulations for tin using our in-house hydrocode FLAG \cite{osti_1663163,osti_2202604} that we compare to the experimental data presented above in Section \ref{sec:tingundata}.
The simulations use the Sesame 2163 equation of state\footnote{While the analytic equations of state outlined in Section \ref{sec:eos} were very useful for the ramp loading calculations implemented in our python research code \cite{pyptkinetics} and discussed in the previous section above, the present shock-loading simulations using LANL hydrocode FLAG use tabulated equations of state from the LANL database `Sesame', see Ref. \cite{Rehn:2023tinrpt} for a description of the tin EOS we use from this database.} \cite{Rehn:2023tinrpt}, scaled versions of the Preston-Tonks-Wallace (PTW) strength model \cite{PTW:2003} for $\beta$ and $\gamma$ tin, and either the Greeff model or the newly implemented Fermi model for the phase transition kinetics.
The rate-dependent Greeff+ model was not considered due to its unphysical hysteresis curve as discussed in the previous section.

For deviatoric strength in both phases of the tin at the high strain rates of shock testing, we follow the work in  Refs. \cite{Nguyen:2024} and \cite{Prime:2026a}.
The PTW strength model parameters were calibrated on constitutive data up to $4000$ s$^{-1}$ for $\beta$-tin in Ref. \cite{Nguyen:2024}.
Lacking a calibration of PTW parameters for $\gamma$-tin (due to the inability to perform simple constitutive testing on the high-pressure phase), Ref. \cite{Prime:2026a} uses the same set of parameters that were calibrated for $\beta$-tin, but use a scale factor for the flow stress, calibrated on shock-driven Richtmyer-Meshkov Instability (RMI) data.
Because the RMI data, at about $10^{7}$s$^{-1}$, extrapolates well beyond the calibration regime, a scale factor on the $\beta$-tin strength was also needed to match the data.

All simulations were carried out with a resolution of 1$\mu$m.
This zone size was determined to be small enough for our purpose by repeating the same simulation with different zone sizes until the results converged.

As pointed out in Section \ref{sec:greeff}, the two fitting parameters of the phenomenological Greeff Model \cite{Greeff:2016,GreefKPT}, are in fact quite correlated, making it hard to fit to real phase transitions.
To highlight this problem, we compare shock data on tin (dashed lines showing the free surface velocity as a function of time) with FLAG simulations (solid lines) in Figure \ref{fig:fitgreeff}.

With the Fermi model developed above in Section \ref{sec:fermimodel}, we already get a better fit shown in Figure \ref{fig:fitFermi}, even if we ignore any rate dependence of the model parameters (i.e. $\gamma_b=0=\gamma_w$).
With this set of only two parameters we cannot ``spread out'' the curves more, i.e. we cannot move the cyan curve to the left without simultaneously also moving the purple curve to the left and thus, cannot optimize the fit across different shock strengths or strain rates.
Upon including the rate pressure dependence, we are able to improve the fit further, as shown in Figures \ref{fig:fitFermimod2} and \ref{fig:fitFermimod2b}.
In addition to making the two model parameters pressure rate dependent, we also adjusted the strength scaling factors within the bounds given by Ref. \cite{Prime:2026a} to further improve the fits.

While the tweaked scale factor especially improved the onset of the first plateau for all curves, the additional rate dependence improves the `spread' of the curves so that the lower impact velocity ones (green, blue, red, black) can be moved more to the right while keeping the high velocity (cyan) one almost in place.
Although the improvement due to the rate dependence in the present fit is only moderate, we remind the reader of the comparison to high-rate data on iron shown in Figure \ref{fig:compareSmith} which served as the main reason to introduce the rate dependence of the Fermi model.

\begin{table}[ht]
\centering
\begin{tabular}{l|c|r||c|r}
 & $\beta$-tin / $\gamma$-tin & figure(s) & $\alpha$-iron / $\epsilon$-iron & figure \\\hline
Fermi Model &&& \\[0.5em]
$w$ [J/kg] &  \shortstack{5e3\\4e3} & \shortstack{\ref{fig:volfracTin}\\\ref{fig:fitFermi}--\ref{fig:fitFermimod2b}} &  5e3 & \ref{fig:compareSmith}\\[0.5em]
$b$ [J/kg]  & \shortstack{5e2\\3e2} & \shortstack{\ref{fig:volfracTin}\\\ref{fig:fitFermi}--\ref{fig:fitFermimod2b}} & 5e2 & \ref{fig:compareSmith}\\[0.5em]
$\gamma_w$ [$\mu$s] & \shortstack{0.2\\9} & \shortstack{\ref{fig:volfracTin}\\\ref{fig:fitFermimod2}--\ref{fig:fitFermimod2b}} & 0.3 & \ref{fig:compareSmith}\\[0.5em]
$\gamma_b$ [$\mu$s] & \shortstack{0.2\\9} & \shortstack{\ref{fig:volfracTin}\\\ref{fig:fitFermimod2}--\ref{fig:fitFermimod2b}} & 0.5 & \ref{fig:compareSmith}\\[0.5em]
PTW scale & \shortstack{0.65 / 0.1\\0.53 / 0.2} & \shortstack{\ref{fig:fitFermi}\\\ref{fig:fitFermimod2}--\ref{fig:fitFermimod2b}} & -  \\\hline
Greeff (+) Model &&&\\[0.5em]
$\nu_G$ [$\mu$s$^{-1}$] &  \shortstack{0.1\\0.05} & \shortstack{\ref{fig:volfracTin}\\\ref{fig:fitgreeff}} & 5e2 & \ref{fig:compareSmith} \\[0.5em]
$C_2$ [J/kg]  & \shortstack{1e3\\2e3} & \shortstack{\ref{fig:volfracTin}\\\ref{fig:fitgreeff}} & 3e3 & \ref{fig:compareSmith} \\[0.5em]
PTW scale & 0.65 / 0.1 & \ref{fig:fitgreeff} & -  
\\[0.5em]
$\gamma_{\nu G}$ [$\mu$s] &  2e-1 & \ref{fig:volfracTin} & 3e-1 & \ref{fig:compareSmith}\\
$\gamma_{C2}$ [$\mu$s]  & 2e-1 & \ref{fig:volfracTin} & 1e-1 & \ref{fig:compareSmith}
\\\hline
\end{tabular}
\caption{Summary of model parameters discussed in this section for the Fermi model and the Greeff model (with and without rate dependence), referencing the figures that use them.
Since the purpose of Fig. \ref{fig:volfracTin} was to highlight the model's capabilities with regards to the hysteresis even at low strain rates, the values of the model parameters used in that figure are different from the ones calibrated to the high strain rate data.}
\label{tab:modelparams}
\end{table}

When the impact velocity is large enough such that the phase transition is overdriven, the experimental data show a vertical single shock rise (dashed lines in Figure \ref{fig:fitFermimod2b}).
This limiting case is not well captured by the Greeff model as well (solid line in the left panel of Figure \ref{fig:fitFermimod2b}), as it exhibits a pronounced kink.
While a small kink remains even with the rate-dependent Fermi model, the agreement with the data is much better (solid line in the right panel of Fig. \ref{fig:fitFermimod2b}).

Table \ref{tab:modelparams} summarizes all the model parameter values discussed above for the Fermi model and the Greeff model. References to the figures that use them are included as well for the reader's convenience.
EOS parameters are summarized in Table \ref{tab:tineosparam} and the model parameters for the micro-structure dependent model are collected in Table \ref{tab:modelparams_micro}.

The calibrations of all models described in this work can likely be improved further by using a Bayesian calibration tool.
For this work, which is beyond the scope of the present paper, we plan to use the LANL calibration code \href{https://github.com/lanl/impala}{IMPALA}.

\section{Conclusion}

In this paper, we presented and validated our new phenomenological ``Fermi'' kinetic phase transition (KPT) model.
In particular, we compared its fidelity to two other models, the phenomenological Greeff KPT model as well as the micro-structure dependent physics-based (and thereby computationally slower) model of Ref. \cite{Blaschke:2025a}.
As shown in Figures \ref{fig:compareSmith}--\ref{fig:fitFermimod2b}, the new Fermi KPT model performs better than the Greeff model in all cases considered in this work, and gives comparable results to the micro-structure aware model at a fraction of the computational cost.
In the quasi-static limit the Fermi model even outperforms the micro-structure aware model by capturing the hysteresis curve seen in experiments such as \cite{Giles:1971,Merkel:2020}, as highlighted at the example of tin in Figure \ref{fig:volfracTin}.

The Fermi KPT model depends on only 4 model parameters per phase transition which need to be calibrated.
It has been implemented in our in-house hydrocode FLAG as well as the open source research code PyPTkinetics \cite{pyptkinetics}.
In the case of tin, the Fermi KPT model was calibrated using newly measured gas-gun flyer-plate impact data presented in Section \ref{sec:tingundata}.

\subsection*{Acknowledgments}
\noindent
We thank D. L. Preston for related discussions.
DNB and AEM gratefully acknowledge support from the Equations of State project within the Physics and Engineering Models (PEM) Subprogram element of the Advanced Simulation and Computing (ASC) Program at Los Alamos National Laboratory (LANL). 
AH gratefully acknowledges support from the Materials project within the ASC-PEM Program at LANL.
LANL, an affirmative action/equal opportunity employer, is operated by Triad National Security, LLC, for the National Nuclear Security Administration of the U.S. Department of Energy under contract 89233218NCA000001.

\bibliographystyle{utphys-custom}
\bibliography{PTkin_fermi}

\end{document}